# The Semi-Relativistic Equation via the Shifted-l Expansion Technique

T. Barakat[*]

Near East University, Lefkoşa, P.O. BOX 1033 Mersin 10 - Turkey

## Abstract

The semi-relativistic equation is cast into a second-order Schrödinger-like equation with the inclusion of relativistic corrections up to order $(v/c)^2$. The resulting equation is solved via the Shifted-$l$ expansion technique, which has been recently developed to get eigenvalues of relativistic and non-relativistic wave equations. The Coulomb, Oscillator, and the Coulomb-plus-linear potentials used in $q\bar{q}$ phenomenology are tested. The method gives quite accurate results over a wide range of $r$ and any choice of quantum numbers $n$ and $l$. However, a comparison of the present work with those of Lucha et al. and Nickisch et al. will serve as a test of this approach.

[*]E-mail: zayd@cc.ciu.edu.tr

# 1  Introduction

The problem of solving the semi-relativistic equation (SR) is a very old subject, but it still an important theme in the existing literature. In recent years much attention has been focused on the semi-relativistic equation, because it help in understanding the structure of heavy-light mesons, especially the decay constant of the $B$ meson $f_B$, and the Cabibbo-Kobayaski-Maskawa (CKM) matrix elements $V_{cb}$ and $V_{ub}$.

This spatial wave equation (SR), is the more realistic starting point for the description of bound states consisting of scalar bosons as well as of the spin-averaged spectra of bound states consisting of two fermions of masses $m_1$, $m_2$ and total energy $M$. In the past few years, the (SR) equation has been tested for its bound-state energies numerically and analytically using different techniques [2] and references therein. All these attempts to obtain the energy eigenvalues and corresponding wave functions of bound states from (SR) equation rely very heavily on numerical methods. Here, this work formulates the Shifted-$l$ expansion technique (SLET) in which such awkward difficulties may not arise. Very recently we introduce the Shifted-$l$ expansion technique (SLET) to solve the Schrödinger [3], Dirac [4], and Klein-Gordon equations for some model potentials [5]. SLET simply consists of using $1/\bar{l}$ as an expansion parameter, where $\bar{l} = l - \beta$. $l$ is the angular momentum, and $\beta$ is a suitable shift introduced partly to avoid the trivial case $l = 0$. In addition to that, $\beta$ is chosen so that the next leading term in the energy eigenvalue series vanishes, and therefore improving the convergency of the perturbation series. This choice is physically motivated by requiring agreement between SLET analytical expressions for the eigenvalues and the modified eigenfunctions to reproduce the exact analytical expressions of the harmonic Oscillator and Coulomb potentials [6,7]. Unlike other perturbation methods [8-11], SLET puts no constraints on the coupling constants of the potential or on the quantum numbers involved. Therefore, the main motivation of the present work is to overcome the shortcomings of previous approaches and to formulate an elegant algebraic approach to yield a fairly simple analytic formula which will give a rapidly converging leading-orders of the energy values with good accuracy. It is worth mentioning that SLET not only reproduces sufficiently accurate energy values for coulomb plus linear potential but it also gives exact analytic results for the ground- state energy coulomb problem.

In this spirit, this paper is organized as follows. Section 2, recalls the conversion of th SR equation into an equivalent Schrödinger- type equation, and there the Shifted-$l$ expansion technique (SLET) for this equation with any spherically symmetric potential is introduced. The analytical expressions for SLET's are cast in such a way that allows the reader to use them without proceeding into their derivation. Finally, Sec.3. being devoted to test our technique for Coulomb, Oscillator interactions, and for the realistic Coulomb-plus-linear potential used in $q\bar{q}$ phenomenology, and then we conclude and remark therein.

## 2  SLET for the Semi-Relativistic Equation with Spherically Symmetric Potentials.

The (SR) equation is the combination of relativistic kinematics with some static interaction potential [1],

$$[(p^2 + m_1^2)^{1/2} + (p^2 + m_2^2)^{1/2} + V(r) - M]\,\psi(r) = 0\,. \tag{1}$$

An expansion in the powers of $(v/c)^2$ up to two terms yields

$$\left\{\frac{p^2}{2\mu} - \frac{p^4}{8\nu} + V(r)\right\}\psi_{n\ell}(r) = E_{n\ell}\,\psi_{n\ell}(r)\,, \tag{2}$$

where $E_{n\ell} = M - m_1 - m_2$, $\mu = m_1 m_2/(m_1 + m_2)$ is the reduced mass and $\nu = m_1^3 m_2^3/(m_1^3 + m_2^3)$ is a useful parameter. The first two terms and $(m_1 + m_2)$ follow from the expansion of $(p^2 + m_1^2)^{1/2} + (p^2 + m_2^2)^{1/2}$. In this reduction formalism the spherically symmetric potential $V(r)$ which represents the interaction between the two particles remains unspecified. In the physical literature, the second term in Eq.(2) is treated as perturbation by using trial wave functions [12]. In this work, in order to obtain a Schrödinger-like equation, this term is treated using the reduced Schrödinger equation [13]:

$$p^4 = 4\mu^2\left[M - m_1 - m_2 - V(r)\right]^2\,. \tag{3}$$

Thus, the second order Schrödinger–like equation to order $(v/c)^2$ becomes

$$\left\{\frac{p^2}{2\mu} - \frac{1}{2\eta}[E_{n\ell}^2 + V^2(r) - 2E_{n\ell}V(r)] + V(r)\right\}\psi_{n\ell}(r) = E_{n\ell}\psi_{n\ell}(r), \tag{4}$$

where $\eta = \nu/\mu^2$. Writing the operator $p^2$ in the spherical polar coordinates

$$p^2 = -\frac{d^2}{dr^2} - \frac{2}{r}\frac{d}{dr} + \frac{l(l+1)}{r^2}, \tag{5}$$

and for states of definite orbital angular momentum $l$, define the reduced radial wave function $R_{n\ell}(r)$ by $\psi_{n\ell}(r) = r^{-1}R_{n\ell}(r)Y_{lm}(\theta, \phi)$, and after some manipulation Eq.(4) can be transformed to a Schrödinger-like equation ( in units $\hbar = c = 1$) for the radial wave function $R_{n\ell}(r)$

$$\left[-\frac{1}{2\mu}\frac{d^2}{dr^2} + \frac{l(l+1)}{2\mu r^2} + \gamma(r) + \frac{E_{n\ell}V(r)}{\eta}\right]R_{n\ell}(r) = \left(\frac{E_{n\ell}^2}{2\eta} + E_{n\ell}\right)R_{n\ell}(r), \tag{6}$$

with $\gamma(r) = V(r) - V^2(r)/2\eta$.

If the angular momentum $l$ is shifted through the relation $l = \bar{l} + \beta$, Eq.(6) becomes

$$\left[-\frac{1}{2\mu}\frac{d^2}{dr^2} + \frac{[\bar{l}^2 + \bar{l}(2\beta + 1) + \beta(\beta + 1)]}{2\mu r^2} + \gamma(r) + \frac{E_{n\ell}V(r)}{\eta}\right]R_{n\ell}(r) = \left(\frac{E_{n\ell}^2}{2\eta} + E_{n\ell}\right)R_{n\ell}(r), \tag{7}$$

where $n$ in this paper is the radial quantum number.

The systematic procedure of SLET begins with shifting the origin of the coordinate through the definition

$$x = \bar{l}^{1/2}(r - r_o)/r_o \tag{8}$$

where $r_o$ is currently an arbitrary point to perform Taylor expansions about, with its particular value to be determined below. Expansions about this point yield:

$$\frac{1}{r^2} = \sum_{j=0}^{\infty}(-1)^j \frac{(j+1)}{r_o^2} x^j \bar{l}^{-j/2}, \tag{9}$$

$$V(x(r)) = \sum_{j=0}^{\infty} \frac{d^j V(r_o)}{dr_o^j} \frac{(r_o x)^j}{j!} \bar{l}^{-j/2}, \tag{10}$$

$$\gamma(x(r)) = \sum_{j=0}^{\infty} \frac{d^n V(r_o)}{dr_o^j} \frac{(r_o x)^j}{j!} \bar{l}^{-j/2}. \tag{11}$$

It should be mentioned here that the re-scaled coordinate, Eq.(8), has no effect on the energy eigenvalues, which are coordinate-independent. It just facilitates the calculations of both the energy eigenvalues and wave functions. It is also convenient to expand $E_{n\ell}$ as:

$$E_{n\ell} = \sum_{i=0}^{\infty} E_i \bar{l}^{-i}. \tag{12}$$

Substituting Eqs.(9)-(12) into Eq.(7), and expanding around $x = 0$ in powers of $x$ and $\bar{l}$, one gets;

$$\left[ \frac{-1}{2\mu} \frac{d^2}{dx^2} + \frac{1}{2\mu}(\bar{l} + (2\beta + 1) + \frac{\beta(\beta+1)}{\bar{l}})(1 - \frac{2x}{\bar{l}^{1/2}} + \frac{3x^2}{\bar{l}} - \cdots) \right.$$

$$+ \frac{r_o^2 \bar{l}}{Q}(\gamma(r_o) + \frac{\gamma'(r_o)r_o x}{\bar{l}^{1/2}} + \frac{\gamma''(r_o)r_o^2 x^2}{2\bar{l}} + \frac{\gamma'''(r_o)r_o^3 x^3}{6\bar{l}^{3/2}} + \cdots)$$

$$\left. + \frac{r_o^2 \bar{l}}{\eta Q}(V(r_o) + \frac{V'(r_o)r_o x}{\bar{l}^{1/2}} + \cdots)(E_o + \frac{E_1}{\bar{l}} + \frac{E_2}{\bar{l}^2} + \cdots) \right] \phi_{n\ell}(x)$$

$$= \mathcal{E}_{n\ell} \phi_{n\ell}(x), \tag{13}$$

where the primes of $V(r_o)$, and $\gamma(r_o)$ denotes derivatives with respect to $r_o$,

$$\mathcal{E}_{n\ell} = \frac{r_o^2}{Q}\left[ \bar{l}(E_o + \frac{E_o^2}{2\eta}) + (E_1 + \frac{E_o E_1}{\eta}) + (E_2 + \frac{E_o E_2}{\eta} + \frac{E_1^2}{2\eta})\frac{1}{\bar{l}} \right.$$

$$\left. + (E_3 + \frac{E_o E_3}{\eta} + \frac{E_1 E_2}{\eta})\frac{1}{\bar{l}^2} + \cdots \right], \tag{14}$$

and Q is a constant that scales the potential, and is set, for any specific choice of $\bar{l}$, equal to $\bar{l}^2$ at the end of the calculations. This is because, the angular momentum barrier term goes like $\bar{l}^2$ at large $\bar{l}$, so should the potential $V(r)$. Equation (13) is exactly of the type of Schrödinger-like equation for the one-dimensional anharmonic oscillator and has been investigated in detail for spherically symmetric potentials by Imbo et al [6]:

$$\left[ -\frac{1}{2\mu} \frac{d^2}{dr^2} + \frac{1}{2}\mu\omega^2 x^2 + \epsilon_o + P(x) \right] \chi_{n\ell}(x) = \lambda_{n\ell} \chi_{n\ell}(x). \tag{15}$$

To estimate $P(x)$ and $\epsilon_o$, one has to match the terms of Eqs. (13-14) with the terms in Eq.(15) consequently, the final analytic expression for the energy eigenvalues appropriate to (SR) particle is

$$\mathcal{E}_{n\ell} = \bar{l}\left[\frac{1}{2\mu} + \frac{r_o^2 V(r_o)E_o}{\eta Q} + \frac{r_o^2 \gamma(r_o)}{Q}\right]$$

$$+ \left[\frac{(2\beta+1)}{2\mu} + \frac{r_o^2 V(r_o)E_1}{\eta Q} + (n+\frac{1}{2})w\right]$$

$$+ \frac{1}{\bar{l}}\left[\frac{\beta(\beta+1)}{2\mu} + \frac{r_o^2 V(r_o)E_2}{\eta Q} + \alpha_1\right]$$

$$+ \frac{1}{\bar{l}^2}\left[\frac{r_o^2 V(r_o)E_3}{\eta Q} + \alpha_2\right], \tag{16}$$

where $\alpha_1$ and $\alpha_2$ appearing as a correction to the leading order of the energy expression are defined and listed in the appendix of Ref. [4], and the relevant quantities $\epsilon's$ and $\delta's$ corresponding to (SR) equation are given in the appendix of this paper. A quantitative estimate for the energy terms in Eq.(16) can be obtained by comparing the terms of same order in $\bar{l}$ of Eq.(14). The calculations, and the results are very simple,

$$E_o = V(r_0) - \eta + \sqrt{\eta^2 + \frac{\eta Q}{\mu r_0^2}}, \tag{17}$$

$$E_2 = \frac{Q\alpha_{(1)}}{r_0^2 \left(1 + \frac{E_0 - V(r_0)}{\eta}\right)}, \tag{18}$$

$$E_3 = \frac{Q\alpha_{(2)}}{r_0^2 \left(1 + \frac{E_0 - V(r_0)}{\eta}\right)}. \tag{19}$$

Here, $\beta$ is chosen so that the next contribution to the leading term in the energy eigenvalue series to vanish, i.e., $E_1 = 0$, which implies that

$$\beta = -1/2 - \mu(n+1/2)\omega, \tag{20}$$

where
$$\omega = \frac{1}{\mu}\left[3 + r_0 V''(r_0)/V'(r_0) - \mu r_0^4 V'(r_0)^2/(Q\eta)\right]^{1/2}, \tag{21}$$

and $Q$ satisfies
$$Q = \frac{\mu}{2\eta}\left[r_0^2 V'(r_0)\right]^2 (1+\xi), \tag{22}$$

with
$$\xi = \sqrt{1 + [2\eta/r_0 V'(r_0)]^2}. \tag{23}$$

On the other hand, $r_0$ is chosen to minimize $E_0$, such that,
$$\frac{dE_0}{dr_0} = 0 \quad \text{and} \quad \frac{d^2 E_0}{dr_0^2} > 0, \tag{24}$$

collecting these terms, and carrying out the mathematics, one can get
$$1 + 2\ell + \mu(2n+1)\omega = r_0^2 V'(r_0)\left(\frac{2\mu}{\eta} + \frac{2\xi\mu}{\eta}\right)^{1/2}, \tag{25}$$

which is an explicit equation in $r_0$. Substituting the expressions of Eq.(17), Eq.(18) and Eq.(19) into Eq.(12) immediately gives an expression for the energy eigenvalues, that is
$$E_{n\ell} = E_0 + \frac{\alpha_{(1)}}{r_0^2\left(1 + \frac{E_0 - V(r_0)}{\eta}\right)} + \frac{\alpha_{(2)}}{r_0^2\left(1 + \frac{E_0 - V(r_0)}{\eta}\right)\bar{l}} + O\left[\frac{1}{\bar{l}^2}\right]. \tag{26}$$

It is convenient to summarize the above procedure in the following steps: (a) Calculate $r_o$ from Eq.(25) and substitute it in Eq.(22) to find $Q$ (b) Substitute $Q$ in Eq.(17) to obtain $E_o$. (c) Finally, one can obtain $E_o$ and then calculate $E_{nl}$ from Eq.(26). However, one is not always able to calculate $r_o$ in terms of the potential coupling constants since the analytical expressions become algebraically complicated, although straightforward. Therefore, one has to appeal to numerical computations to find $r_o$ and hence $E_o$.

# 3 Applications, results, and discussion.

In the existing literature, considerable interest has been attached to the analytic solution of the semi-relativistic Coulomb ground-state energy problem. Such ground-state energy have been derived in Ref.[14,15], and it is given as

$$\tilde{E}_n = \frac{2m}{\left[1 + \frac{\alpha^2}{(2n+2)^2}\right]^{1/2}}, \qquad n = 0, 1, 2, ... \tag{27}$$

On the other hand, an analytical upper bound on the ground-state energy level have been obtained by some straightforward variational procedure [16], which reads

$$\hat{E}_n \leq 2m \left[1 - \frac{\alpha^2}{(2n+2)^2}\right]^{1/2}, \qquad n = 0, 1, 2, ... \tag{28}$$

In this work, in order to remove this inadequacy, we propose the Shifted$l$ expansion technique, SLET, to solve the (SR) equation, which not only restores the exact analytic expression for Coulomb ground-state energy problem, but also predicts quite accurate results for the Oscillator, and Coulomb-plus-linear potentials. Therefore, it is worthwhile to illustrate that our formulas Eq.(17), (20), (21), and (22) provide a remarkable accurate and simple analytic expressions to the S-wave (SR) equation energy eigenvalues of spherically symmetric Coulomb potential $V(r) = -\alpha/r$. This test is stringent to this approach. Following the SLET procedure one obtains $\omega = 2/m$, with $m = m_1 = m_2$,

$$Q = (2n+2)^2/4, \tag{29}$$

and

$$r_o = \frac{(2n+2)^2}{2m} \left[\frac{1}{\alpha^2} - \frac{1}{(2n+2)^2}\right]^{1/2}. \tag{30}$$

Substitution of the above expressions into Eq.(17), immediately gives an expression for the S-wave eigenvalues

$$E_o = 2m \left[1 - \frac{\alpha^2}{(2n+2)^2}\right]^{1/2} - 2\,m, \tag{31}$$

hence,

$$M = 2m \left[1 - \frac{\alpha^2}{(2n+2)^2}\right]^{1/2}, \qquad n = 0, 1, 2, ... \tag{32}$$

In Table 1, where the $l = 0$ eigenvalues obtained from SLET are compared with the coulomb analytical result obtained by [14,15], and by some straightforward variational procedure [16]. It is found that, the eigenvalues obtained from Eq.(31) are in good agreement with the results obtained by [14,15], and match very well an upper bound derived by [16].

In a second test, Eq.(26) is solved as stated in Sec. 2, it is shown that the solutions for the Oscillator potential are very well behaved. In Table 2, the convergence of the eigenvalues with increasing $n$, and $l$, with the results obtained in Ref.[2] are quit well.

Finally, the rate of convergence of the eigenvalues obtained from SLET for a realistic Coulomb-plus-linear potential are shown in Table 3 compared to other methods.

As a result, it is observed that, the convergence of the eigenvalues are very rapid as $n$, and $l$ are increasing. Therefore, we can say, this technique is easy to implement, the results are sufficiently accurate for practical purposes, and the SLET equations are cast in away to be used to the unequal mass cases $q\bar{Q}$. Moreover, substantial computation time reduction has been achieved.

Table 1: Energy eigenvalues (in units of GeV) of the $(SR)$ equation, Eq.(31), for the Coulomb potential $V(r) = -\alpha/r$ with $\alpha = 0.25$, and $m_1 = m_2 = m = 1.45$ GeV. The values in round parentheses are those of square root method, whereas, the values in square parentheses are those of integral method Ref. [2].

|  | n=0 | n=1 | n=2 | n=3 | n=4 | n=5 |
|---|---|---|---|---|---|---|
| $\tilde{E}_{no}$ | -0.022394 | -0.005648 | -0.002514 | -0.001415 | -0.000906 | -0.000629 |
| $E_{no}$ | -0.02274 | -0.00566 | -0.002516 | -0.001415 | -0.000906 | -0.000629 |
| $E_{no}$ | (-0.02306) | (-0.00574) | (-0.00254) | (-0.00143) | (-0.000912) | (-0.000635) |
| $E_{no}$ | [-0.02251] | [-0.00556] | [-0.00244] | [-0.00140] | [-0.00085] | [-0.00065] |

Table 2: Energy eigenvalues (in units of GeV) of the $(SR)$ equation, Eq.(26), for the Oscillator potential $V(r) = r^2/2$ with $m_1 = m_2 = m = 1.310$ GeV. The values in round parentheses are those of square root method [2], whereas, the values in square parentheses are those of Miller's method Ref. [2, 17].

|        |          |          | $E_{n\ell}$ (SLET) |          |          |
|--------|----------|----------|----------|----------|----------|
|        | n=0      | n=1      | n=2      | n=3      | n=4      |
| $\ell=0$ | 1.6536   | 3.5048   | 5.1409   | 6.6269   | 8.0049   |
|        | (1.6595) | (3.5280) | (5.1654) | (6.6553) | (8.0395) |
| Miller | [1.6595] | [3.5280] | [5.1654] | [6.6553] | [8.0394] |
| $\ell=1$ | 2.6609   | 4.3719   | 5.9218   | 7.3484   | 8.6823   |
|        | (2.6663) | (4.3932) | (5.9441) | (7.3737) | (8.7125) |
| Miller | [2.6663] | [4.3932] | [5.9441] | [7.3737] | [8.7124] |
| $\ell=2$ | 3.6066   | 5.2086   | 6.6844   | 8.0577   | 9.3508   |
|        | (3.6110) | (5.2268) | (6.7039) | (8.0796) | (9.3766) |
| Miller | [3.6110] | [5.2268] | [6.7039] | [8.0796] | [9.3765] |

Table 3: Energy eigenvalues (in units of GeV) of the $(SR)$ equation, Eq.(26), for $V(r) = -\alpha/r + \beta r$ with $\alpha = 0.25$, $\beta = 0.18\ GeV^2$, and $m_1 = m_2 = m = 1.45$ GeV. The values in round parentheses are those of square root method Ref. [2].

|        |          |          | $E_{n\ell}$ (SLET) |          |          |
|--------|----------|----------|----------|----------|----------|
|        | n=0      | n=1      | n=2      | n=3      | n=4      |
| $\ell=0$ | 0.4930   | 1.0069   | 1.3988   | 1.7323   | 2.0295   |
|        | (0.4924) | (1.0022) | (1.3925) | (1.7252) | (2.0205) |
| $\ell=1$ | 0.8342   | 1.2484   | 1.5971   | 1.9053   | 2.1855   |
|        | (0.8345) | (1.2481) | (1.5960) | (1.9033) | (2.1793) |
| $\ell=2$ | 1.0958   | 1.4600   | 1.7796   | 2.0685   | 2.3345   |
|        | (1.0962) | (1.4601) | (1.7797) | (2.0687) | (2.3346) |

## Appendix

The definition of $\varepsilon'_j s$ and $\delta'_i s$ for $\alpha_1$ and $\alpha_2$ in Ref.[4] are given here which are applicable to the $(SR)$ equation

$$\bar{\varepsilon}_i = \frac{\varepsilon_i}{(2\mu\omega)^{i/2}}, \quad i = 1, 2, 3, 4. \quad \bar{\delta}_j = \frac{\delta_j}{(2\mu\omega)^{j/2}}, \quad j = 1, 2, 3, 4, 5, 6, \tag{33}$$

where

$$\varepsilon_1 = -(2\beta + 1)/\mu \ , \quad \varepsilon_2 = 3(2\beta + 1)/2/\mu, \tag{34}$$

$$\varepsilon_3 = -2/\mu + \frac{r_o^5}{6Q}\left[\gamma'''(r_o) + V'''(r_o)E_o/\eta\right], \tag{35}$$

$$\varepsilon_4 = 5/2\mu + \frac{r_o^6}{24Q}\left[\gamma''''(r_o) + V''''(r_o)E_o/\eta\right], \tag{36}$$

$$\delta_1 = -\beta(\beta + 1)/\mu + \frac{r_o^3 V'(r_o)E_2}{Q\eta} \tag{37}$$

$$\delta_2 = 3\beta(\beta + 1)/2\mu + \frac{r_o^4 V''(r_o)E_2}{2Q\eta}, \tag{38}$$

$$\delta_3 = -2(2\beta + 1)/\mu \ , \quad \delta_4 = 5(2\beta + 1)/2\mu, \tag{39}$$

$$\delta_5 = -3/\mu + \frac{r_o^7}{120Q}\left[\gamma'''''(r_o) + V'''''(r_o)E_o/\eta\right], \tag{40}$$

$$\delta_6 = 7/2\mu + \frac{r_o^8}{720Q}\left[\gamma''''''(r_o) + V''''''(r_o)E_o/\eta\right]. \tag{41}$$

The terms including $E_1$ have been dropped from the above expressions since $E_1$=0.